\documentclass[12pt]{article}
\usepackage{epsf}
\usepackage[utf8]{inputenc}
\usepackage[english]{babel}

\begin{document}

\title{
{\bf Impact of the $f$-Reggeon exchanges on the observables of the single diffractive dissociation of nucleon at ultrahigh energies}}
\author{A.A. Godizov\thanks{E-mail: anton.godizov@gmail.com}\\
{\small {\it A.A. Logunov Institute for High Energy Physics}},\\ {\small {\it NRC ``Kurchatov Institute'', 142281 Protvino, Russia}}}
\date{}
\maketitle

\vskip-1.0cm

\begin{abstract}
Single diffractive dissociation (SDD) of nucleon in high-energy proton-proton and proton-antiproton collisions is considered in terms of a simple two-Reggeon model with nonlinear 
Regge trajectories. It is demonstrated that the $f$-Reggeon impact on the corresponding cross-sections is not negligible up to the LHC energies. As well, it is shown that the account 
of the $f$-Reggeon exchanges allows to describe the elastic diffractive scattering (EDS) and SDD of nucleons at ultrahigh energies in the framework of a unified phenomenological 
scheme. The predictive value of the proposed model is verified.
\end{abstract}

\section*{1. Introduction}

Diffractive phenomena play a very important role in high-energy dynamics of hadrons. For example, in proton-proton collisions at the LHC energies, the fraction of EDS events in the 
total number of events is $\sim$ 25\%. Another important diffractive reaction is the SDD of proton, $p+p\to p+X$, where $X$ is the produced hadronic state of mass 
${\rm M}_{\rm X}$, while the energy fraction lost by the surviving proton does not exceed 0.05. More than 10\% of the $p\;p$ collision events at the LHC are the SDD events. 

The slopes of the experimentally observed angular distributions point to the fact that the high-energy diffractive scattering of hadrons is related to comparatively large transverse 
distances between interacting particles, from 0.1 to 10 fm. As a consequence, direct applying of perturbative QCD is not relevant, and, hence, we have to exploit phenomenological 
models to estimate the diffractive scattering observables. 

The most natural theoretical framework for high-energy hadron diffraction physics is Regge theory \cite{collins}, where the amplitudes of diffraction reactions are interpreted in 
terms of Reggeon exchanges between hadrons and, as well, of interaction between Reggeons themselves. The main problem which emerges in the Regge phenomenology of high-energy 
hadron-hadron scattering is the fact that we do not know, {\it a priori}, the analytic behavior of the main dynamical characteristics of Reggeons (namely, their Regge trajectories and 
their couplings to hadrons), and, thus, we have to consider these functions to be unknown. Nonetheless, despite of this uncertainty, Regge theory still keeps to be the most promising 
approach to high-energy hadron diffraction due to the universality of Regge trajectories and Reggeon couplings to nucleon (they must be the same in different reactions).

It is demonstrated in \cite{godizov} that it is possible to describe the diffraction pattern of nucleon-nucleon EDS at the collision energy values higher than 10 GeV in the framework 
of a very simple Regge-eikonal approximation where the eikonal is just a sum of two single-Reggeon exchange terms. The first term is related to the so-called ``soft Pomeron'' (SP), a 
supercritical Reggeon which entirely dominates in the nucleon-nucleon EDS at ultrahigh energies. The second term is related to the $f$-Reggeon (FR), a secondary Reggeon associated 
with mesons $f_2(1270)$, $f_4(2050)$, and $f_6(2510)$. Despite of its simplicity, the two-Reggeon model presented in \cite{godizov} possesses a significant predictive value.

The FR contribution to the EDS eikonal turns out to be negligible at the SPS, Tevatron, and LHC energies. The same physical pattern takes place in alternative Regge models: at 
collision energies higher than 500 GeV, the ignoring of secondary Reggeon exchange terms in the EDS eikonal does not lead to any noticeable changes in the diffraction pattern. As a 
consequence, the FR influence is, as well, completely ignored in the most of Regge models which try to describe available data on the proton SDD at ultrahigh energies.

The aim of this paper is to demonstrate that, in fact, the FR impact on the proton SDD is not negligible up to the LHC energies and, also, to show that the account of the FR 
exchanges is absolutely necessary for satisfactory description of available data on the corresponding angular distributions and other observables. That can be done if we combine the 
Regge models for the EDS and SDD of nucleons in the framework of a unified phenomenological scheme with the universal Regge trajectories and couplings to nucleon of both the SP and 
the FR.

\section*{2. Two-Reggeon model with nonlinear Regge trajectories for the SDD of nucleons at ultrahigh energies}

\begin{figure}[ht]
\vskip-0.3cm
\begin{center}
\epsfxsize=8cm\epsfysize=6cm\epsffile{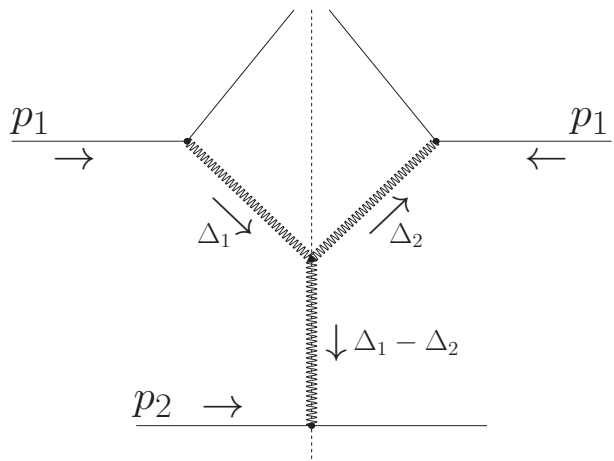}
\end{center}
\vskip-0.7cm
\caption{The diagram for the PPP-interaction bare amplitude.}
\label{triple}
\end{figure}
\vskip 0.3cm

Usually, the reactions of high-energy EDS and SDD are considered in terms of the Mandelstam variables $s=(p_1+p_2)^2$ and $t=(p_1-p_1')^2$, where $p_1$ and $p_2$ are the 4-momenta of 
the incoming particles and $p_1'$ is the 4-momentum of the scattered nucleon. However, if we take account of the in-channel absorption (the importance of such absorptive corrections 
in SDD is discussed in detail in \cite{maor} and \cite{martynov}), then it is more convenient to use such a quantity as the transverse transferred momentum $\vec\Delta_\perp$ related 
to $t$ by means of the relation $t=-(\vec\Delta_\perp^2+\xi^2m_p^2)/(1-\xi)+O(1/s)\approx-\vec\Delta_\perp^2$, where $m_p$ is the proton mass and 
$\xi=({\rm M}_{\rm X}^2-m_p^2)/s\ll1$ is the energy fraction lost by the surviving particle. Below we omit the subscript ``$\perp$'' and treat all the nonscalar variables as 
vectors orthogonal to the beam axis.

The vast majority of Regge models for the SDD of nucleon at ultrahigh energies (higher, than 500 GeV) exploit the approximation of triple-Pomeron interaction (the 
PPP-approximation) which is grounded on the Mueller generalized optical theorem \cite{mueller} and is valid in the asymptotic kinematic regime 
$\sqrt{s}\gg {\rm M}_{\rm X}\gg \{m_p,|\vec\Delta|\}$. The bare PPP-interaction amplitude (see Fig. \ref{triple}) has the following form \cite{collins,godizov2}: 
$$
T_{\rm PPP}(s\,,\,{\rm M}_{\rm X}\,,\,\vec\Delta_1\,,\,\vec\Delta_2) = \frac{1}{s}\left(i+\tan\frac{\pi(\alpha_{\rm SP}(-\vec\Delta_1^2)-1)}{2}\right)
\left(-i+\tan\frac{\pi(\alpha_{\rm SP}(-\vec\Delta_2^2)-1)}{2}\right)\times
$$
$$
\times\;g^{(p)}_{\rm SP}(-\vec\Delta_1^2)\;g^{(p)}_{\rm SP}(-\vec\Delta_2^2)\;g^{(p)}_{\rm SP}(-|\vec\Delta_1-\vec\Delta_2|^2)\;
g_{\rm PPP}(-\vec\Delta_1^2\,,\,-\vec\Delta_2^2\,,\,-|\vec\Delta_1-\vec\Delta_2|^2)\;\times
$$
\vskip 0.1cm
\begin{equation}
\label{tripamp}
\times\;\pi^3\alpha'_{\rm SP}(-\vec\Delta_1^2)\,\alpha'_{\rm SP}(-\vec\Delta_2^2)\,\alpha'_{\rm SP}(-|\vec\Delta_1-\vec\Delta_2|^2)\;\times
\end{equation}
\vskip 0.1cm
$$
\times\;\left(\frac{s}{2s_0}\right)^{\alpha_{\rm SP}(-\vec\Delta_1^2)\,+\,\alpha_{\rm SP}(-\vec\Delta_2^2)}
\left(\frac{{\rm M}_{\rm X}^{\,2}}{2s_0}\right)^{\alpha_{\rm SP}(-|\vec\Delta_1-\vec\Delta_2|^2)\,-\,\alpha_{\rm SP}(-\vec\Delta_1^2)\,-\,\alpha_{\rm SP}(-\vec\Delta_2^2)}\;\;\sim
$$
\vskip 0.1cm
$$
\sim\;\;\frac{1}{s}\left(\frac{1}{\xi}\right)^{\alpha_{\rm SP}(-\vec\Delta_1^2)\,+\,\alpha_{\rm SP}(-\vec\Delta_2^2)}
\left(\frac{{\rm M}_{\rm X}^{\,2}}{2s_0}\right)^{\alpha_{\rm SP}(-|\vec\Delta_1-\vec\Delta_2|^2)}\,,
$$
where $s_0=1$ GeV$^2$, $\alpha_{\rm SP}(t)$ is the SP Regge trajectory, $g^{(p)}_{\rm SP}(t)$ is the SP coupling to nucleon, and the symmetric function $g_{\rm PPP}(t_1,t_2,t_3)$ is 
related to the PPP-interaction vertex. Such factors as $\pi\alpha'_{\rm SP}$ and $2^{-\alpha_{\rm SP}}$ are singled out within the Regge residue for the same reasons as in the case of 
EDS \cite{godizov}.

From the phenomenological standpoint, this approximation seems very attractive due to its simplicity. However, it was shown in \cite{godizov2} that the contribution (\ref{tripamp}) 
is not enough, and the secondary Reggeon impact should be taken into account. Inclusion of the FR into consideration implies a certain modification of the 
PPP-approximation:
\vskip -0.3cm
\begin{equation}
\label{tripampF}
T_{\rm PPP}(s\,,\,{\rm M}_{\rm X}\,,\,\vec\Delta_1\,,\,\vec\Delta_2)\,\to\,T_{bare}(s\,,\,{\rm M}_{\rm X}\,,\,\vec\Delta_1\,,\,\vec\Delta_2)\,=
\,T_{\rm PPP}(s\,,\,{\rm M}_{\rm X}\,,\,\vec\Delta_1\,,\,\vec\Delta_2)\;+
$$
\vskip -0.3cm
$$
+\;\;T_{\rm PPF}(s\,,\,{\rm M}_{\rm X}\,,\,\vec\Delta_1\,,\,\vec\Delta_2)\,+\,T_{{\rm PFF}}(s\,,\,{\rm M}_{\rm X}\,,\,\vec\Delta_1\,,\,\vec\Delta_2)\,,
\end{equation}
\vskip 0.1cm
\noindent where $T_{\rm PPF}(s\,,\,{\rm M}_{\rm X}\,,\,\vec\Delta_1\,,\,\vec\Delta_2)$ is a sum of those three contributions which are obtained from (\ref{tripamp}) via 
replacement of one of the three SPs to the FR, while $T_{\rm PFF}(s\,,\,{\rm M}_{\rm X}\,,\,\vec\Delta_1\,,\,\vec\Delta_2)$ is a sum of those three terms which can be obtained by the 
corresponding replacements of two SPs.

If we introduce the Fourier-image of this function, 
\begin{equation}
\label{fourier}
T_{bare}(s\,,\,{\rm M}_{\rm X}\,,\,\vec b_1\,,\,\vec b_2)\equiv\frac{1}{(16\pi^2s)^2}
\int d^2\vec\Delta_1 d^2\vec\Delta_2\;e^{i\vec\Delta_1\vec b_1}\;T_{bare}(s\,,\,{\rm M}_{\rm X}\,,\,\vec\Delta_1\,,\,\vec\Delta_2)\;e^{-i\vec\Delta_2\vec b_2}\,,
\end{equation}
then the invariant one-particle inclusive cross-section (with account of the absorption in the in-channel) can be represented as \cite{eden}
\begin{equation}
\label{diffcross}
16\pi^2s\frac{d^2\sigma_{\rm SDD}}{dt d{\rm M}_{\rm X}^2} = (4s)^2\int d^2\vec b_1d^2\vec b_2\;e^{-i\vec\Delta\vec b_1}e^{i\vec\Delta\vec b_2}
\left[e^{i\delta(s,b_1)}\,T_{bare}(s\,,\,{\rm M}_{\rm X}\,,\,\vec b_1\,,\,\vec b_2)\,e^{-i\delta^*(s,b_2)}\right]\,,
\end{equation}
where $\delta(s,b)$ is the EDS eikonal in the impact parameter representation \cite{godizov}:
$$
\delta(s,b) = \frac{1}{16\pi^2 s}\int d^2\vec\Delta\;e^{i\vec\Delta\vec b}\;\Omega(s,-\vec\Delta^2) = \frac{1}{16\pi s}\int_0^{\infty}d(\vec\Delta^2)\,
J_0(b\,|\vec\Delta|)\,\Omega(s,-\vec\Delta^2)= 
$$
\begin{equation}
\label{eik}
= \frac{1}{16\pi s}\int_0^{\infty}d(-t)\,J_0(b\sqrt{-t})\left\{\left(i+\tan\frac{\pi(\alpha_{\rm SP}(t)-1)}{2}\right)
g^{(p)^2}_{\rm SP}(t)\;\pi\alpha'_{\rm SP}(t)\left(\frac{s}{2s_0}\right)^{\alpha_{\rm SP}(t)}+\right.
\end{equation}
\vskip -0.1cm
$$
\left.+\,\left(i+\tan\frac{\pi(\alpha_{\rm FR}(t)-1)}{2}\right)g^{(p)^2}_{\rm FR}(t)\;\pi\alpha'_{\rm FR}(t)\left(\frac{s}{2s_0}\right)^{\alpha_{\rm FR}(t)}\right\}\,.
$$

To make quantitative predictions for the SDD observables, we need to fix the model degrees of freedom, namely, the unknown functions $\alpha_{\rm SP}(t)$, $g^{(p)}_{\rm SP}(t)$, 
$\alpha_{\rm FR}(t)$, $g^{(p)}_{\rm FR}(t)$, $g_{\rm PPP}(t_1,t_2,t_3)$, $g_{\rm PPF}(t_1,t_2,t_3)$, and $g_{\rm PFF}(t_1,t_2,t_3)$. 

Any Reggeon is related to a certain family of hadron resonances, and, hence, its Regge trajectory and couplings to various hadrons are universal in different hadron 
reactions. Therefore, for description of the SDD observables, the parametrizations for the SP and FR Regge trajectories and for their couplings to nucleon should be the same as in 
the case of EDS 
\cite{godizov}: 
\begin{equation}
\label{pomeron}
\alpha_{\rm SP}(t) = 1+\frac{\alpha_{\rm SP}(0)-1}{1-\frac{t}{\tau_{\rm SP}}}\;,\;\;\;\;\alpha_{\rm FR}(t) = \frac{\alpha_{\rm FR}(0)}{1-\frac{t}{\tau_{\rm FR}}}\;,
\end{equation}
$$
g^{(p)}_{\rm SP}(t)=\frac{g^{(p)}_{\rm SP}(0)}{(1-a_1\,t)^2}\;,\;\;\;\;g^{(p)}_{\rm FR}(t)=\frac{g^{(p)}_{\rm FR}(0)}{(1-a_2\,t)^{3/2}}\;,
$$
\vskip 0.3cm
\noindent where the free parameters take on the values presented in Table \ref{tab1}.
\vskip 0.2cm
\begin{table}[ht]
\begin{center}
\begin{tabular}{|l|l|}
\hline
\bf Parameter          & \bf Value        \\
\hline
$\alpha_{\rm SP}(0)-1$ & 0.114            \\
$\tau_{\rm SP}$        & 0.552 GeV$^2$    \\
$g^{(p)}_{\rm SP}(0)$  & 13.1 GeV         \\
$a_1$                  & 0.276 GeV$^{-2}$ \\
$\alpha_{\rm FR}(0)$   & 0.61             \\
$\tau_{\rm FR}$        & 1.54 GeV$^2$     \\
$g^{(p)}_{\rm FR}(0)$  & 18.2 GeV         \\
$a_2$                  & 0.47 GeV$^{-2}$  \\
\hline
\end{tabular}
\end{center}
\vskip -0.2cm
\caption{The parameter values for (\ref{pomeron}) obtained via fitting to the proton-proton EDS data (the details can be found in \cite{godizov}).}
\label{tab1}
\end{table}
\vskip 0.2cm

The choice of parametrization (\ref{pomeron}) for $\alpha_{\rm SP}(t)$ and $\alpha_{\rm FR}(t)$ is specified by the asymptotic behavior of the SP and FR Regge trajectories which 
follows from QCD \cite{kearney,kwiecinski}:\footnote{It should be noted that the asymptotic behavior of the parametrization (\ref{pomeron}) for $\alpha_{\rm FR}(t)$ is different from 
the Kwiecinski solution \cite{kwiecinski} obtained within the framework of perturbative QCD in the logarithmic approximation. However, in the proposed model with the parameter values 
from Table \ref{tab1} the relative contribution of the FR exchange term into the EDS eikonal in the region $\sqrt{-t}>$ 1.5 GeV is suppressed by the factor $\sim[s/(2s_0)]^\epsilon$ 
($\epsilon>0.75$) in comparison with the SP exchange term. Therefore, the true analytic asymptotics of $\alpha_{\rm FR}(t)$ is not important for quantitative description of the EDS 
data at high energies.} 
\begin{equation}
\label{asyP}
\lim_{t\to-\infty}\alpha_{\rm SP}(t)=1\,,\;\;\;\;\lim_{t\to-\infty}\alpha_{\rm FR}(t)=0\,.
\end{equation}
In their turn, in the corresponding limit, the parametrizations of the SP and FR couplings to nucleon should obey the so-called Quark Counting Rules \cite{matveev}:
\begin{equation}
\label{asypom2}
\lim_{t\to -\infty}g^{(p)}_{\rm SP}(t) = O(|t|^{-2})\,,\;\;\lim_{t\to -\infty}g^{(p)}_{\rm FR}(t) = O(|t|^{-3/2})\,.
\end{equation}
Thus, the set of functions (\ref{pomeron}) is the simplest approximation to the SP and FR Regge trajectories and their couplings to nucleon which explicitly satisfies the asymptotic 
relations (\ref{asyP}) and (\ref{asypom2}) following from QCD. However, one should keep in mind that the true Regge trajectories and Reggeon form-factors have much more complicated 
analytic structure (with branching\linebreak points, {\it etc.}). Therefore, the exploited test functions are intended to approximate $\alpha_{\rm SP}(t)$, $\alpha_{\rm FR}(t)$, 
$g^{(p)}_{\rm SP}(t)$, and $g^{(p)}_{\rm FR}(t)$ just in the region of negative values of their argument. These parametrizations are in no way suitable for the region $t>0$.

Concerning the triple-Reggeon couplings, in this paper we restrict ourselves by the simplest possible approximation: 
$$
g_{\rm PPP}\left(-\vec\Delta_1^2\,,\,-\vec\Delta_2^2\,,\,-|\vec\Delta_1-\vec\Delta_2|^2\right)\approx g_{\rm PPP}(0,\,0,\,0)\equiv g_{\rm PPP}\,,
$$
\vskip -0.5cm
\begin{equation}
\label{vertex3}
g_{\rm PPF}\left(-\vec\Delta_1^2\,,\,-\vec\Delta_2^2\,,\,-|\vec\Delta_1-\vec\Delta_2|^2\right)\approx g_{\rm PPF}(0,\,0,\,0)\equiv g_{\rm PPF}\,,
\end{equation}
\vskip -0.4cm
$$
g_{\rm PFF}\left(-\vec\Delta_1^2\,,\,-\vec\Delta_2^2\,,\,-|\vec\Delta_1-\vec\Delta_2|^2\right)\approx g_{\rm PFF}(0,\,0,\,0)\equiv g_{\rm PFF}\,.
$$
The assumption (\ref{vertex3}) makes the proposed model very stiff because, in this case, all the observables depend on three constants $g_{\rm PPP}$, $g_{\rm PPF}$, and 
$g_{\rm PFF}$ linearly\footnote{A case of quite different dynamical structure of the Regge residues is discussed in \cite{ryutin}.}, while the free parameters related to 
$\alpha_{\rm SP}(t)$, $\alpha_{\rm FR}(t)$, $g^{(p)}_{\rm SP}(t)$, and $g^{(p)}_{\rm FR}(t)$ are fixed by fitting to the proton-proton EDS angular distributions. Hence, the model 
itself can be easily discriminated (or confirmed) via comparison with available experimental data on the proton SDD, and its applicability range can be determined reliably. In 
what follows, we will see that the considered simple phenomenological approximation is valid in a wide kinematic region.

\section*{3. Model predictions {\it versus} available experimental data}

\begin{figure}[ht]
\vskip -0.5cm
\epsfxsize=8.2cm\epsfysize=8.2cm\epsffile{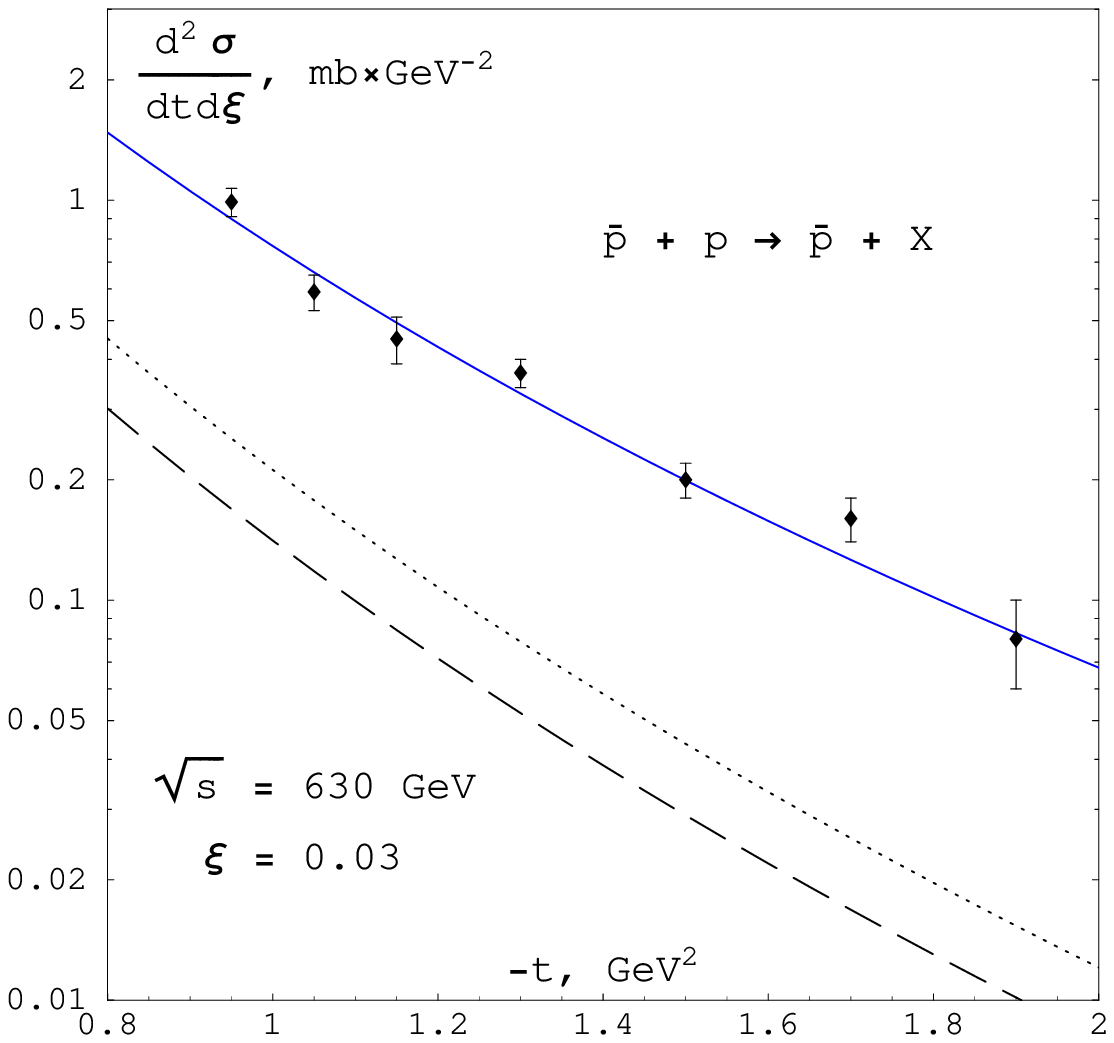}
\vskip -8.15cm
\hskip 8.75cm
\epsfxsize=8.15cm\epsfysize=8.15cm\epsffile{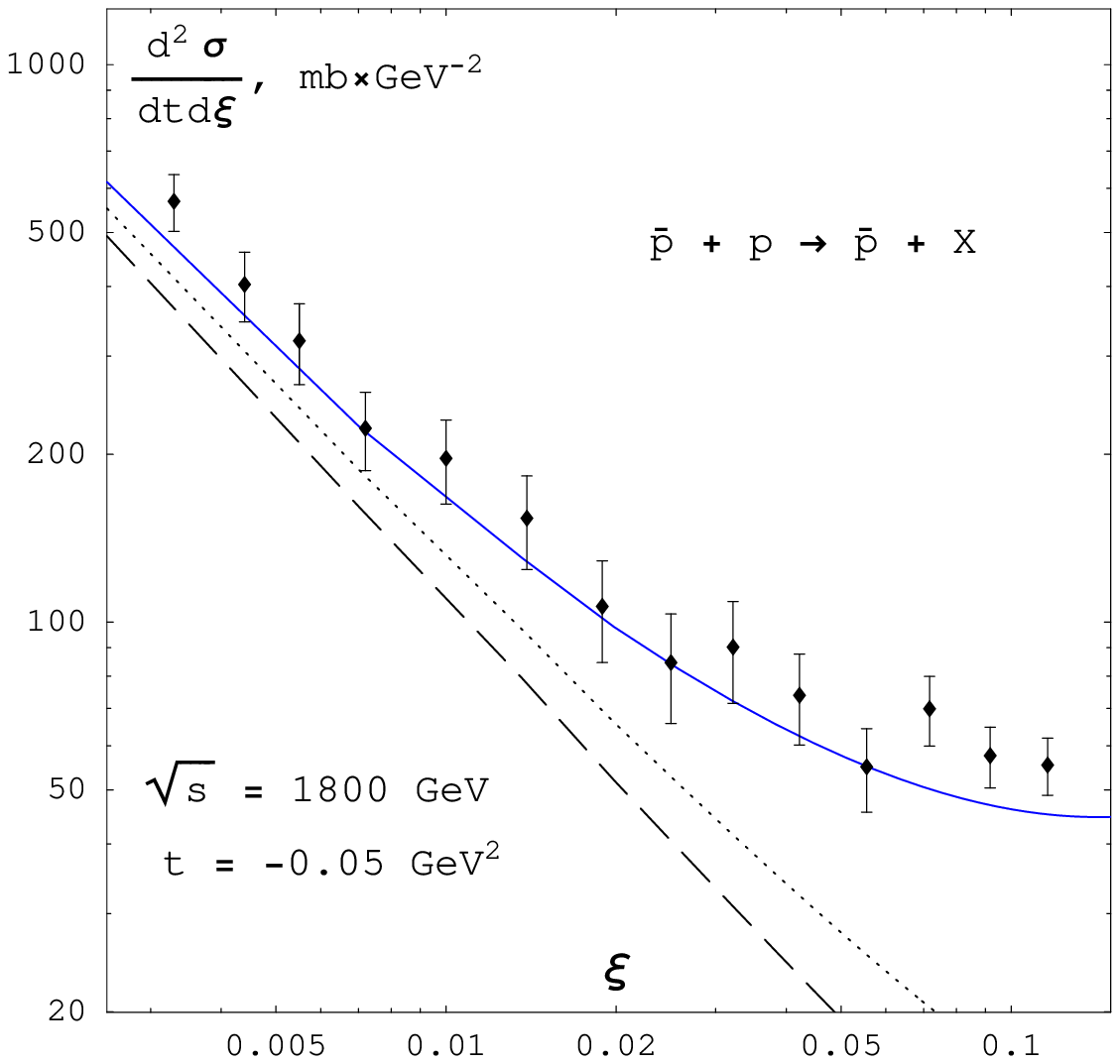}
\vskip -0.2cm
\caption{The two-Reggeon model description of the doubly differential cross-section of the proton SDD in $\bar p\,p$ collisions (the data are taken from \cite{UA8} and \cite{montanha}, 
correspondingly). The dashed (dotted) lines correspond to the contribution of the PPP (PPP $+$ PPF) interaction.}
\label{diffxit}
\end{figure}
\vskip 0.2cm

As we intend not only to describe a finite set of the SDD data, but, also, to check the predictive value of the proposed model, so we need, first, to choose some subset of the total 
considered set of data and to fit the values of $g_{\rm PPP}$, $g_{\rm PPF}$, and $g_{\rm PFF}$ to this dataset. Second, we must use these values to compare the model predictions 
with those data which were not included into the fitting procedure. The chosen dataset includes the SDD integrated cross-sections at the collision energies 546 GeV \cite{UA4}, 
0.9, 2.76, and 7 TeV \cite{ALICE}, 1.8 TeV \cite{E710,E710a}, and 8 TeV \cite{ATLAS}, as well as the UA8 data on the $t$-distribution at $\sqrt{s} =$ 630 GeV and 
$\xi =$ 0.03 \cite{UA8}. No other data were used in the fitting procedure.

\begin{figure}[ht]
\vskip -0.2cm
\epsfxsize=8.05cm\epsfysize=8.05cm\epsffile{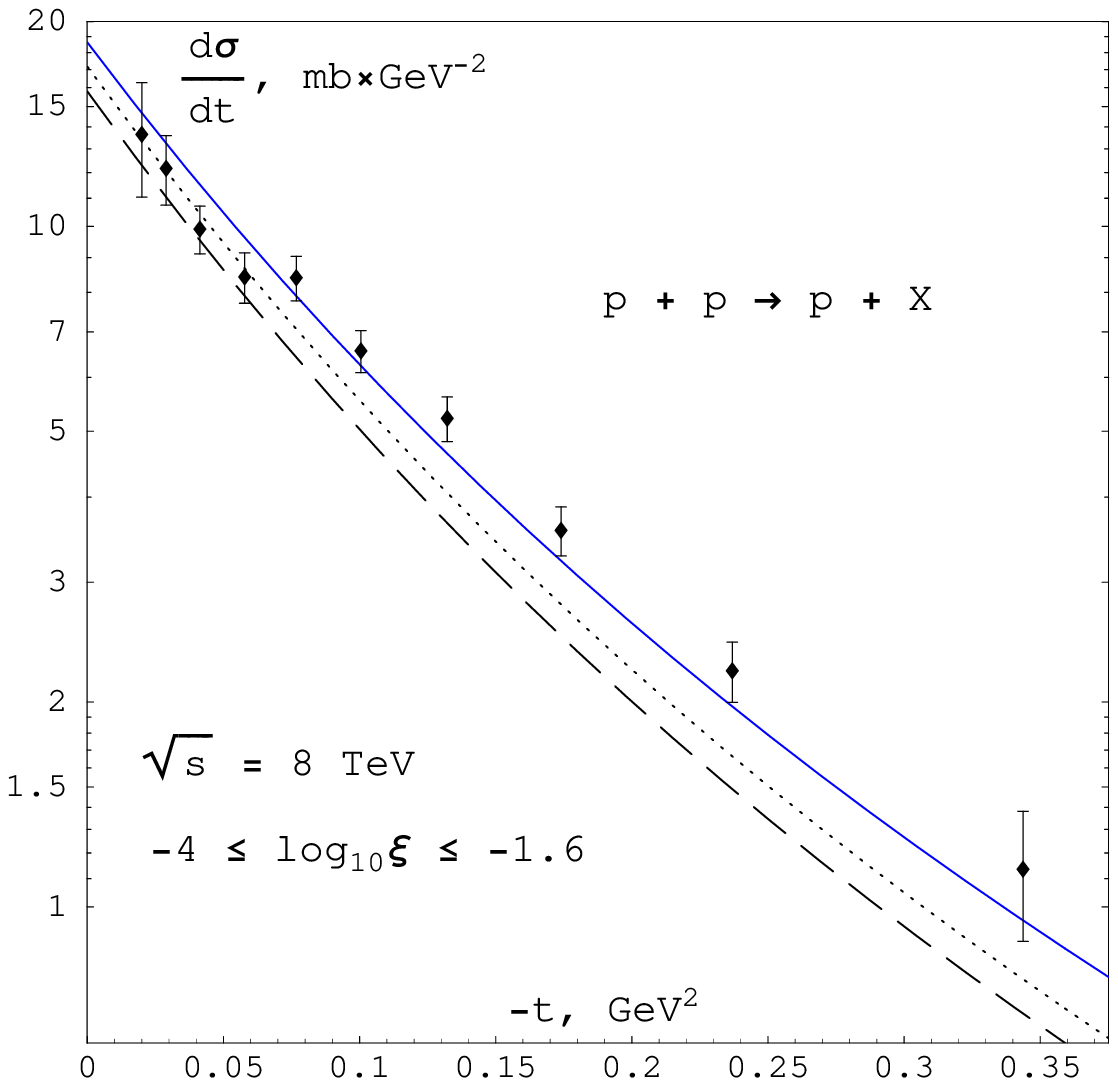}
\vskip -8.1cm
\hskip 8.75cm
\epsfxsize=8.25cm\epsfysize=8.25cm\epsffile{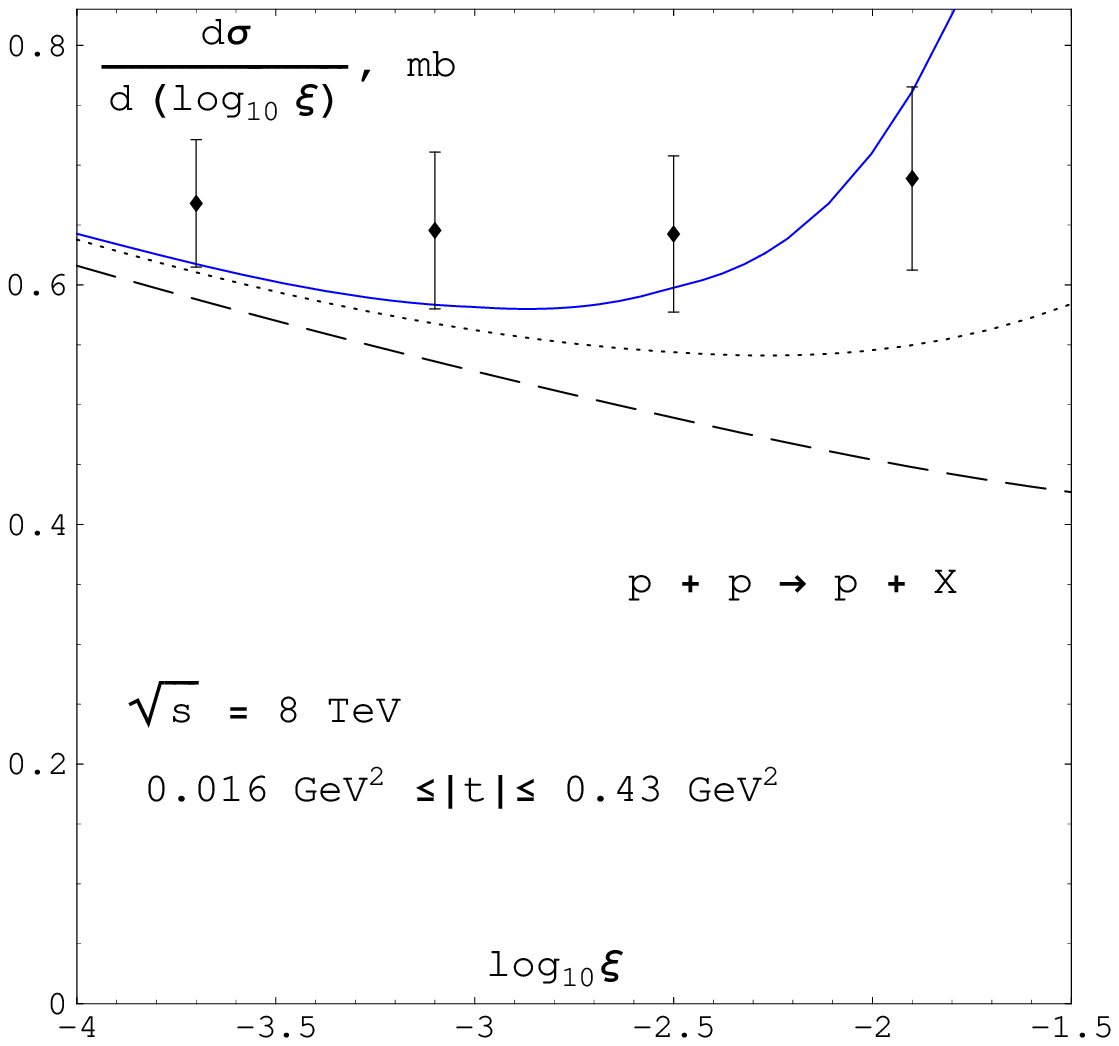}
\vskip -0.2cm
\caption{The two-Reggeon model predictions for the $t$- and $\xi$-distributions in the proton SDD at $\sqrt{s}=8$ TeV (the data are taken from \cite{ATLAS}). The dashed (dotted) lines 
correspond to the contribution of the PPP (PPP $+$ PPF) interaction.}
\label{difft}
\end{figure}

\begin{figure}[ht]
\vskip -0.2cm
\begin{center}
\epsfxsize=8.15cm\epsfysize=8.15cm\epsffile{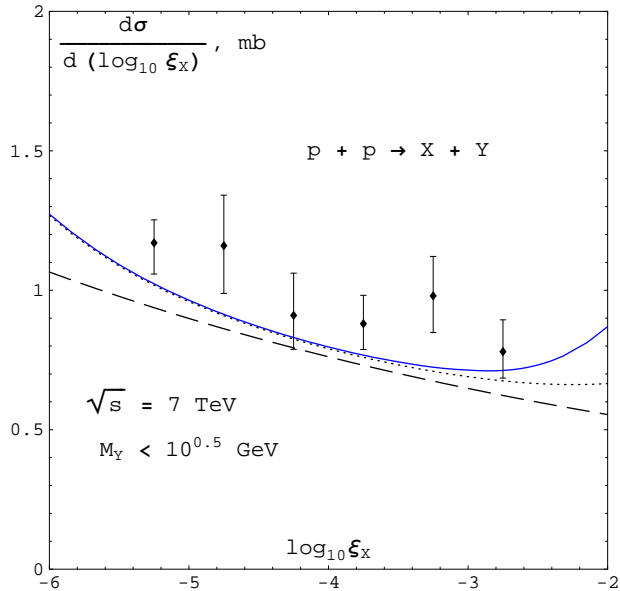}
\end{center}
\vskip -0.4cm
\caption{The two-Reggeon model predictions for the $\xi$-distribution in the proton SDD at $\sqrt{s}=7$ TeV (the data are taken from \cite{CMS}). The dashed (dotted) line corresponds 
to the contribution of the PPP (PPP $+$ PPF) interaction.}
\label{diffxi}
\end{figure}

The fitting results are presented in Fig. \ref{diffxit} (the left picture) and Table \ref{integr}. The obtained values of $g_{\rm PPP}$, $g_{\rm PPF}$, and $g_{\rm PFF}$ are 
\begin{equation}
\label{vertex4}
g_{\rm PPP}\approx 0.5\;{\rm GeV}\,,\;\;\;\;g_{\rm PPF}\approx 0.2\;{\rm GeV}\,,\;\;\;\;g_{\rm PFF}\approx 1.3\;{\rm GeV}\,.
\end{equation}
\vskip -0.3cm
\begin{table}[ht]
\begin{center}
\begin{tabular}{|l|l|l|l|}
\hline
\bf Experiment     & \bf Kinematic range  &  $\sigma^{\rm exp}_{\rm SDD}(s)$, mb & $\sigma^{\rm model}_{\rm SDD}(s)$, mb \\
\hline
UA4   \cite{UA4}   & $\sqrt{s}=546$  GeV,    $m_p+m_{\pi^0}<{\rm M}_{\rm X}<\sqrt{0.05\,s}$ &   9.4  $\pm$ 0.7           &  9.3    \\
ALICE \cite{ALICE} & $\sqrt{s}=0.9$  TeV,    $m_p+m_{\pi^0}<{\rm M}_{\rm X}<200$ GeV        & $11.2^{+1.6}_{-2.1}$       & 10.8    \\
E-710 \cite{E710}  & $\sqrt{s}=1.8$  TeV,    $m_p+m_{\pi^0}<{\rm M}_{\rm X}<\sqrt{0.05\,s}$ &  11.7  $\pm$ 2.3           & 13.3    \\
E-710 \cite{E710a} & $\sqrt{s}=1.8$  TeV     $\sqrt{2}$ GeV$<M_{\rm X}<\sqrt{0.05\,s}$      &   8.1  $\pm$ 1.7           &  9.8    \\
ATLAS \cite{ATLAS} & $\sqrt{s}=8$    TeV,    $80$ GeV$<{\rm M}_{\rm X}<1270$ GeV            &  $3.18  \pm  0.26$         &  3.1    \\
ALICE \cite{ALICE} & $\sqrt{s}=2.76$ TeV,    $m_p+m_{\pi^0}<{\rm M}_{\rm X}<200$ GeV        & $12.2^{+3.9}_{-5.3}$       & 13.2    \\
ALICE \cite{ALICE} & $\sqrt{s}=7$    TeV,    $m_p+m_{\pi^0}<{\rm M}_{\rm X}<200$ GeV        & $14.9^{+3.4}_{-5.9}$       & 16.5    \\
\hline
\end{tabular}
\caption{The two-Reggeon model description of available data on the SDD integrated cross-section 
$\sigma_{\rm SDD}=2\int\int\frac{d^2\sigma_{\rm SDD}}{dt\,dM_{\rm X}^2}\,dt\,dM_{\rm X}^2$.}
\label{integr}
\end{center}
\end{table}
\vskip -0.2cm
The model predictions for the SDD differential cross-sections are presented in Figs. \ref{diffxit} (the right picture), \ref{difft}, and \ref{diffxi}. All the figures accentuate the 
importance of the PPF and PFF interaction contributions (see the difference between the solid, dashed, and dotted lines). 
\vskip 0.2cm

The model has successfully predicted the $\xi$-dependence of $\frac{d^2\sigma_{\rm SDD}}{dt\,d\xi}$ measured at $\sqrt{s} =$ 1.8 TeV and $t = - 0.05$ GeV$^2$ \cite{montanha} (the right 
picture in Fig. \ref{diffxit}) and, as well, the $t$-slope of the angular distribution produced by ATLAS Collaboration \cite{ATLAS} in the region of low $t$ (the left picture in Fig. 
\ref{difft}). However, one can observe a noticeable underestimation of the $\xi$-distribution presented by CMS Collaboration \cite{CMS} (Fig. \ref{diffxi}). This deviation of the 
model curve from the CMS data is due to the fact that the corresponding dataset is contaminated by a significant fraction (about 10-15\%) of the double diffractive dissociation 
events. Regarding the ATLAS data \cite{ATLAS} (where the double diffractive events are reliably excluded), such a divergence does not take place (see the right picture in 
Fig. \ref{difft}). Hence, we can conclude that, despite of its stiffness, the proposed model demonstrates a significant predictive value. Nonetheless, the reliability of the 
considered approximation is closely related to the problem of determination of its applicability range. Let us discuss this subject in more details.

\section*{4. Discussion}

From the practical standpoint, the question about the applicability range of any phenomenological model is the most important. Exploiting the approximation (\ref{tripampF}), 
we ignore all the terms which contain higher powers of $g_{\rm PPP}$, $g_{\rm PPF}$, and $g_{\rm PFF}$. Such contributions to the SDD observables unavoidably emerge in Reggeon Field 
Theory. Neglecting these terms in our phenomenological analysis is justified {\it a posteriori} due to the relative smallness of the parameters $g_{\rm PPP}$, $g_{\rm PPF}$, and 
$g_{\rm PFF}$ in comparison with the SP and FR couplings to nucleon (see (\ref{vertex4}) and Table \ref{tab1}). Therefore, we could assume that these terms are inessential up to the 
LHC energies, and the verified predictive value of the proposed model confirms this assumption. However, one should keep in mind that some of these contributions grow with energy 
faster than simple triple-Reggeon interaction terms. As a consequence, in the range $\sqrt{s}\gg 10$ TeV they may start to play an essential role and, hence, should be taken into 
account. Otherwise, in the region of asymptotically high energies we will face with the so-called Finkelstein-Kajantie (FK) problem, when, in the proposed approximation, the SDD 
integrated cross-section grows with energy faster than $\sim\ln^2s$ (for detailed discussion of the FK problem, see \cite{martynov2} and references therein).

Thus, the considered simple approximation is valid in the limited range of the collision energy values. Nonetheless, the unified phenomenological scheme, which includes the model for 
the nucleon-nucleon EDS \cite{godizov} as well as the proposed model for the nucleon SDD, pretends to be a reliable tool for description of both the reactions in the 
considered kinematic ranges. Certainly, it must be compared with the alternative approaches exploiting the notion of Reggeon and intended for combined description of these processes 
(the key feature is usage of the same Regge trajectories and Reggeon form-factors for both the reactions). The most known EDS models \cite{donnachie1}, \cite{martynov3}, and 
\cite{selyugin}, at their current stage of development, do not provide such a description. However, two approaches exist which try to do it. The first one is the so-called Additive 
Quark Model (AQM) \cite{shuvaev} where baryons are treated as systems of three spatially separated constituent quarks. The amplitudes of diffractive processes in this model are 
described in terms of the SP exchanges only. At present, the AQM did not reach any more or less satisfactory description of available SDD datasets. The second one is the well-known 
KMR model \cite{khoze} which takes account of the FR impact on the SDD observables and provides a much better description of the SDD cross-sections in comparison with the AQM model, 
though some noticeable deviations of the model curves from the CDF data on $\frac{d^2\sigma_{\rm SDD}}{dt\,d\xi}$ \cite{montanha} can be observed in the range\linebreak 
0.003 $<\xi<$ 0.02 (see Fig. 8 in \cite{khoze}). In whole, the kinematic range of validity of the KMR model in both the EDS and SDD of nucleons is much smaller than of the proposed 
two-Reggeon model, and such a state of affairs has certain objective reasons.

The main feature of the considered phenomenological scheme which distinguishes it from other Regge models is the usage of essentially nonlinear approximations to Regge trajectories 
in the region of negative values of their argument, while the vast majority of Regge models for the high-energy EDS and SDD of nucleons exploit linear Regge trajectories. Such an 
essential nonlinearity of $\alpha_{\rm SP}(t)$ and $\alpha_{\rm FR}(t)$ in the region of low negative $t$ follows from the fact that these functions should satisfy the asymptotic 
relations (\ref{asyP}) which, in their turn, follow from QCD. As we ascertained above, explicit account of (\ref{asyP}) in parametrizations for $\alpha_{\rm SP}(t)$ and 
$\alpha_{\rm FR}(t)$ significantly simplifies construction of models suitable for combined description of the high-energy EDS and SDD of nucleons in wide kinematic ranges.

\subsection*{4.1. Impact of the Odderon and other secondary Reggeon exchanges}

One of the most important questions is about the impact of Reggeons $\omega$, $\rho$, and $a$ on the observables of the nucleon SDD in the considered kinematic range. Some arguments 
exist why the contributions of the ${\rm P}\omega\omega$, ${\rm P}\rho\rho$, and ${\rm P} a a$ interaction terms into the nucleon SDD cross-sections are subdominant with respect to 
the above-considered {\rm PFF} interaction term.

In \cite{godizov}, the secondary Reggeon exchanges are treated on the basis of the dual approximation of QCD \cite{rossi}, wherein two types of the secondary Reggeon exchange 
contributions emerge in the nucleon-nucleon EDS eikonal. Type 1 contains the exchange by valence quarks, while\linebreak type 2 does not imply any exchanges by valence quarks 
(see Fig. 2 in \cite{godizov}). The contributions of type 2 are related to the $f$-Reggeon only, because for other secondary Reggeons such terms are suppressed due to the approximate 
conservation of isospin and $G$-parity. In the $p\,p$ scattering, 
the imaginary part of the $f$-Reggeon exchange contribution of type 1 
annihilates the imaginary part of the $\omega$-Reggeon exchange contribution (the same takes place for Reggeons $\rho$ and $a$). Hence, the $f$-Reggeon contribution of type 2 
is the only one which is crucial for the $p\,p$ scattering (except the small vicinity of the diffraction dip at relatively low energies), and just this contribution is taken 
into account in the proposed in \cite{godizov} two-Reggeon approximation for the $p\,p$ elastic scattering.

Next, let us consider the $\bar p\,p$ EDS angular distribution at $\sqrt{s}=$ 13.8 GeV (see Fig. 5), where the secondary Reggeon contributions of type 1 to the imaginary part of the 
eikonal have the same sign and are expected to have an essential impact on the value of the differential cross-section. We see that this impact is not so crucial, though it is visible 
in the region of low $t$-values. In its turn, the splitting of the $p\,p$ and $\bar p\,p$ angular distributions at the ISR energies in the region of the Coulomb-nuclear interference 
is, mainly, due to electromagnetic interaction (see Fig. 9 in \cite{godizov}). As well, if we look at Fig. 6 in \cite{godizov} for the integrated cross-sections, we will not see 
much difference between the solid curve (the two-Reggeon approximation) and the data for both the $p\,p$ and $\bar p\,p$ scattering, while the deviations of the dotted line (the 
approximation entirely ignoring the $f$-Reggeon exchanges) from the data at $\sqrt{s}<$ 15 GeV are huge. Consequently, the combined secondary Reggeon contribution of type 1 to both the 
$p\,p$ and $\bar p\,p$ EDS amplitudes is much more subdominant than the $f$-Reggeon contribution of type 2, and, thus, the secondary Reggeon contributions of type 1 can be ignored, in 
the leading approximation. The only reason for such a pattern is that the coupling of secondary Reggeon to nucleon through multi-gluon exchanges is essentially stronger than the 
coupling with the exchange by valence quarks. For the same reason one can expect that the contributions of the ${\rm P}\omega\omega$, ${\rm P}\rho\rho$, and ${\rm P} a a$ interaction 
terms into the SDD cross-sections in the considered kinematic ranges are also subdominant with respect to the {\rm PFF} interaction term and, thus, they can be neglected, in the 
roughest approximation. The verified predictive value of the proposed two-Reggeon model argues in favor of such an assumption.

\begin{figure}[ht]
\begin{center}
\vskip -0.2cm
\epsfxsize=10cm\epsfysize=10cm\epsffile{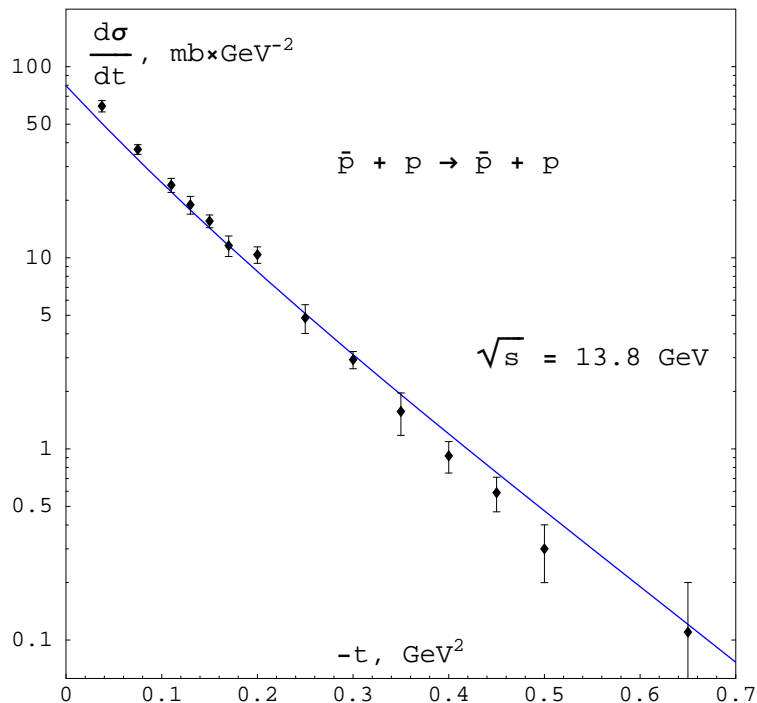}
\end{center}
\vskip -0.6cm
\label{diffppbar}
\caption{The $\bar p\,p$ differential cross-section at $\sqrt{s}=$ 13.8 GeV. The solid line is the prediction of the two-Reggeon eikonal model \cite{godizov} with the SP and FR Regge 
trajectories and the corresponding vertex functions fitted to the $p\,p$ EDS data in the kinematic region \{9.8 GeV $\le\sqrt{s}\le$ 200 GeV, 0.1 GeV $\le\sqrt{-t}\le$ 1.5 GeV\}. The 
data are taken from \cite{ayres}.}
\end{figure}
\vskip 0.3cm

A few words should be said about possible influence of the Odderon, a $C$-odd counterpartner of the SP. The Odderon is responsible for the splitting between the $p\,p$ and $\bar p\,p$ 
EDS angular distributions at ultrahigh energies. In the recent paper \cite{d0totem}, the data on the $p\,p$ EDS at\linebreak $\sqrt{s}=$ 2.76 TeV \cite{totem} extrapolated to 
$\sqrt{s}=$ 1.96 TeV were compared with the data on the $\bar p\,p$ EDS at the same energy \cite{d0} (direct data on the $p\,p$ EDS at $\sqrt{s}=$ 1.96 TeV do not exist, unfortunately). 
The visible splitting between these two datasets in the vicinity of the diffraction dip confirms the presence of the Odderon contribution (see Fig. 4 in \cite{d0totem}). However, the 
absence of such a splitting outside of this vicinity points to the fact that the Odderon contribution to the EDS eikonal is subdominant in comparison with the SP exchange term. 
The most possible reason of this subdominance is the smallness of the Odderon coupling to nucleon. Therefore, in the leading approximation, we may ignore the Odderon impact on the 
observables of the nucleon SDD as well.

\section*{5. Conclusions}

In view of the aforesaid, we conclude:
\begin{itemize}
\item In spite of its roughness, the proposed two-Reggeon model with nonlinear Regge trajectories for the nucleon SDD is valid in a wide kinematic region: 
\{0.5 TeV $<\sqrt{s}\le$ 8 TeV,\linebreak $-t<2$ GeV$^2$, $M_{\rm X}>3$ GeV, $\xi\le$ 0.03\}. Together with the two-Reggeon eikonal approximation for the high-energy EDS of nucleons 
\cite{godizov} it allows to consider the reactions $p\,(\bar p) + p\to p\,(\bar p) + p$ and $p\,(\bar p) + p\to p\,(\bar p) + X$ in the framework of a unified phenomenological scheme.
\item Contrary to the case of EDS, the impact of the $f$-Reggeon exchanges on the SDD observables is not negligible up to the LHC energies.
\end{itemize}

\end{document}